\documentclass[english,english,twocolumn,reprint,prxquantum,nofootinbib,longbibliography,superscriptaddress,dvipsnames]{revtex4-2}

\usepackage[utf8]{inputenc}
\usepackage[T1]{fontenc}
\usepackage{braket}

\usepackage{amsmath}
\usepackage{amssymb}
\usepackage{mathptmx}
\usepackage{bm}
\usepackage{textgreek}

\usepackage{graphicx}
\usepackage{dcolumn}
\newcommand{\figref}[1]{Fig.~\ref{#1}}

\usepackage{etoolbox}
\usepackage[separate-uncertainty=true,multi-part-units=single]{siunitx}

\usepackage{color}
\usepackage{soul}
\usepackage{ulem}
\definecolor{darkblue}{rgb}{0,0,0.7}
\definecolor{darkgreen}{rgb}{0.2,0.95,0.2}
\definecolor{darkred}{rgb}{.7,0,0}
\definecolor{purple}{rgb}{0.7,0,0.9}
\definecolor{orange}{rgb}{1,0.5,0}
\definecolor{grey}{rgb}{0.6,0.6,0.6}
\definecolor{lightpink}{rgb}{1,0.7,0.75}
\definecolor{pink}{rgb}{1,0.4,0.58}

\begin{document}

\title{An ultracompact dilution refrigerator for fast quantum device characterization}

\author{Cl\'ement Geffroy}
\thanks{These two authors contributed equally.}
\affiliation{Univ. Grenoble Alpes, CNRS, Grenoble INP, Institut N\'eel, 38402 Grenoble, France}

\author{Dorian Nicolas}
\thanks{These two authors contributed equally.}
\affiliation{Univ. Grenoble Alpes, CNRS, Grenoble INP, Institut N\'eel, 38402 Grenoble, France}

\author{Eric Eyraud}
\affiliation{Univ. Grenoble Alpes, CNRS, Grenoble INP, Institut N\'eel, 38402 Grenoble, France}

\author{Shelender Kumar}
\affiliation{Univ. Grenoble Alpes, CNRS, Grenoble INP, Institut N\'eel, 38402 Grenoble, France}

\author{Supriya Mandal}
\affiliation{Univ. Grenoble Alpes, CNRS, Grenoble INP, Institut N\'eel, 38402 Grenoble, France}

\author{Julien Jarreau}
\affiliation{Univ. Grenoble Alpes, CNRS, Grenoble INP, Institut N\'eel, 38402 Grenoble, France}

\author{Laura Kowalski}
\affiliation{Univ. Grenoble Alpes, CNRS, Grenoble INP, Institut N\'eel, 38402 Grenoble, France}

\author{Laurent Del-Rey}
\affiliation{Univ. Grenoble Alpes, CNRS, Grenoble INP, Institut N\'eel, 38402 Grenoble, France}

\author{Didier Dufeu}
\affiliation{Univ. Grenoble Alpes, CNRS, Grenoble INP, Institut N\'eel, 38402 Grenoble, France}

\author{Nicolas Roch}
\affiliation{Univ. Grenoble Alpes, CNRS, Grenoble INP, Institut N\'eel, 38402 Grenoble, France}

\author{Wolfgang Wernsdorfer}
\affiliation{Physikalisches Institut, Karlsruhe Institute of Technology, Wolfgang-Gaede-Str. 1, Karlsruhe, D-76131, Germany.}

\author{Quentin Ficheux}
\email{quentin.ficheux@neel.cnrs.fr}
\affiliation{Univ. Grenoble Alpes, CNRS, Grenoble INP, Institut N\'eel, 38402 Grenoble, France}

\author{Matias Urdampilleta	}
\email{matias.urdampilleta@neel.cnrs.fr}
\affiliation{Univ. Grenoble Alpes, CNRS, Grenoble INP, Institut N\'eel, 38402 Grenoble, France}

\date{\today}

\begin{abstract}
Rapid thermal cycling is a central bottleneck in the development of superconducting quantum devices: conventional dilution refrigerators require cooldowns of a day or more and substantial cryogenic infrastructure, which throttles the fabricate--measure--redesign loop. We present an ultracompact dilution refrigerator---\qty{3}{\kilo\gram} in mass and \qty{100}{\milli\meter} in diameter---that completes a full cooldown–warm-up cycle to a base temperature of \qty{70}{\milli\kelvin} in \qty{1.2}{\hour} when unloaded, and in \qty{2.1}{\hour} when fully equipped with the microwave wiring required for qubit measurements, while delivering \qty{20}{\micro\watt} of cooling power at \qty{100}{\milli\kelvin}. We validate the platform through a complete characterization of a two-fluxonium device: we extract the full circuit Hamiltonian by two-tone spectroscopy, measure energy-relaxation and coherence times, and benchmark single-qubit control. Although the relaxation time is limited by the base temperature of the system, we reach a single-qubit gate fidelity of up to \qty{99}{\percent}, at the coherence limit set by our operating temperature. These results demonstrate that compact, fast-cycling dilution refrigeration can support state-of-the-art quantum-device characterization without sacrificing measurement quality, offering a practical route to high-throughput quantum-hardware development.
\end{abstract}

\maketitle

\section{Introduction}

The development of quantum hardware relies on rapid cycles of device fabrication, cooldown, measurement, and redesign. In superconducting quantum circuits, progress in coherence, control fidelity, and architectural scalability has been tightly coupled to advances in cryogenic measurement infrastructure ~\cite{Mohseni2024}. The ability to reproducibly characterize devices at millikelvin temperatures remains a central requirement for validating new materials, circuit designs, and fabrication processes.

Over the past decade, large-scale dilution refrigerators have enabled dramatic improvements in superconducting qubit performance, delivering long coherence times and high-fidelity operations~\cite{ZU2022103390}. At the same time, the growing complexity of quantum circuits and the exploration of novel device concepts have created an increasing need for rapid experimental iteration. As a result, measurement throughput has emerged as a critical factor limiting the pace of quantum hardware development.

Conventional dilution refrigerators, however, are not optimized for such iterative workflows. These systems typically require cooldown times of a day or more and involve substantial operational overhead, including high electrical power consumption and complex cryogenic infrastructure. While they provide excellent stability for long-duration experiments, these characteristics constrain the speed at which new devices can be tested and refined.

To address these limitations, recent efforts have explored alternative cryogenic platforms aimed at improving accessibility, reducing system size, and accelerating experimental turnaround such as bottom-loader systems ~\cite{batey2014rapid}. Compact and cryogen-free dilution refrigerators have been developed to reduce footprint and simplify operation~\cite{10.1063/1.4972249, Guan_2024, Krinner2019}, while other approaches have focused on fast-cooldown strategies and lightweight cryogenic designs~\cite{10.1063/5.0139825, 10384560, Barber2024}. In parallel, dedicated studies have investigated the compatibility of such platforms with microwave measurements and quantum device characterization~\cite{PhysRevApplied.18.L041002}. Collectively, these works represent important progress toward more agile cryogenic infrastructures.

A key challenge nonetheless remains. Existing compact or fast-cycling systems generally involve trade-offs between cooldown speed, base temperature, and measurement performance, and only a limited number of demonstrations have combined rapid, resource-efficient cryogenic operation with state-of-the-art superconducting qubit measurements on the same platform~\cite{10384560, Barber2024, PhysRevApplied.18.L041002}. The extent to which compact cryogenic systems can support high-coherence quantum devices without degrading their performance therefore remains an open question.

Here, we address this challenge with an ultracompact dilution refrigerator designed specifically for rapid, resource-efficient quantum device testing. By minimizing thermal mass and streamlining the cryogenic architecture, the system achieves markedly reduced cooldown times while maintaining stable millikelvin operation, and is engineered to support high-quality microwave measurements compatible with superconducting qubit characterization. We demonstrate its performance by fully characterizing a two-fluxonium device integrated directly within the cryostat: we extract the complete circuit Hamiltonian, measure coherence, and benchmark single-qubit control. The resulting metrics are comparable to those obtained in conventional large-scale dilution refrigerators at the same temperature and with equivalent shielding and wiring, showing that fast and compact cryogenic operation can be achieved without compromising device characterization. These results establish compact dilution refrigeration as a practical pathway toward high-throughput quantum device characterization, addressing a key bottleneck in quantum-hardware development~\cite{10384560, Barber2024, PhysRevApplied.18.L041002} and opening the way to more scalable and adaptative quantum engineering workflows.

\section{Cryostat Architecture}

\begin{figure}[t]
    \centering
    \includegraphics[width=1.0\linewidth]{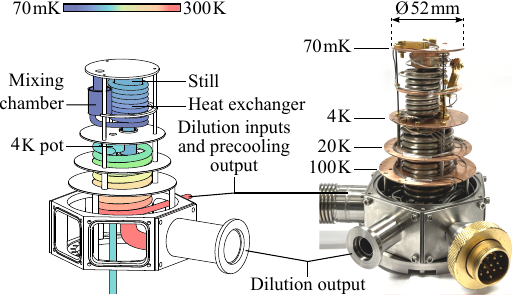}
    \caption{
        Compact inverted dilution cryostat.
        Schematic and photograph of the cryogenic platform highlighting the inverted geometry, with the mixing chamber located at the top and the warm stages at the bottom.
        The system features a \qty{52}{\milli\meter}-diameter top flange within a \qty{100}{\milli\meter}-diameter vacuum can and provides accessible experimental volumes at both the \qty{4}{\kelvin} and millikelvin stages.
        The main cryogenic circuits, including the $^4$He pot, precooling loop, and dilution unit, are indicated.
    }
    \label{fig:1_sionludiXS}
\end{figure}

The cryostat adopts an inverted geometry in which the mixing chamber plate sits at the very top of the structure, giving the user direct access to the coldest stage and thereby simplifying sample mounting and modification of experimental components. As shown in \figref{fig:1_sionludiXS}, the warmest stage is located at the bottom (the \qty{300}{\kelvin} interface), and the intermediate plates (\qty{100}{\kelvin}, \qty{20}{\kelvin}, and \qty{4}{\kelvin}) follow in ascending order. Although the cryostat as a whole is inverted and the mixing chamber is the topmost stage, the still remains above the mixing chamber. A ring-shaped cold plate at approximately \qty{200}{\milli\kelvin} provides an additional thermalization point below the mixing-chamber plate.

The entire system weighs only \qty{3}{\kilo\gram}, including radiation shields, and its \qty{100}{\milli\meter} outer diameter gives it an exceptionally small footprint. As a result, it can be deployed alongside existing experimental setups without dedicated infrastructure, or integrated into confined environments such as the access ports of high-field magnets. It is also well suited to modular cryogenic architectures, in which densely integrated cryogenic subsystems interconnected by quantum links are arranged in close proximity~\cite{Storz2023, Photonic2024, Mirhosseini2020, Schar2026, Magnard2020}. Despite the reduced size, the design preserves a functional experimental volume of \qty{150}{\centi\meter\cubed} at the \qty{4}{\kelvin} stage and \qty{680}{\centi\meter\cubed} at the mixing-chamber stage, sufficient to accommodate auxiliary cryogenic RF components such as attenuators, filters, and amplifiers.

Three cryogenic circuits operate in parallel. A standard $^4$He circuit cools the \qty{4}{\kelvin} stage through a helium pot; a $^3$He--$^4$He dilution circuit provides millikelvin operation; and a dedicated precooling circuit accelerates the initial thermalization of the upper stages from room temperature to \qty{4}{\kelvin}. The integration of these circuits is optimized to minimize dead volumes and improve thermal efficiency, and particular care is taken to preserve robust thermal anchoring of all components despite the constrained geometry. At base temperature, the mixing chamber delivers a cooling power of \qty{20}{\micro\watt} at \qty{100}{\milli\kelvin}, which, as shown below, is sufficient to support a complete qubit characterization.

\subsection{Cryogenic operation}

The cooling protocol is designed to minimize the total experimental cycle time while ensuring reproducible operation. In the initial step, the system is precooled from room temperature to \qty{4}{\kelvin} by circulating the $^3$He--$^4$He mixture through the helium pot, relying on efficient heat exchange between the gas mixture and the cryostat structure. In the unloaded configuration, thermalization to \qty{4}{\kelvin} is reached within approximately \qty{15}{\minute}.

Once the system reaches \qty{4}{\kelvin}, operation switches to dilution mode. The \qty{5}{\liter} mixture, of which only \qty{1}{\liter} is $^3$He, is condensed, leading to phase separation in the dilution unit; cooling is then driven by the enthalpy of mixing between $^3$He and $^4$He. The mixing-chamber temperature falls from \qty{4}{\kelvin} to approximately \qty{70}{\milli\kelvin} within \qty{25}{\minute}, and the base temperature is stabilized by continuous circulation of the mixture. The warm-up phase, initiated by stopping circulation and letting the system equilibrate with its environment, typically requires about \qty{30}{\minute}. The total cycle time in the unloaded configuration is therefore approximately \qty{1.2}{\hour}.

\begin{figure}[t]
    \centering
    \includegraphics[width=1\linewidth]{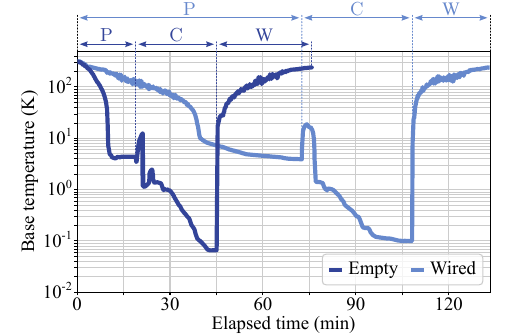}
    \caption{
        Cryogenic cycle and temperature evolution during operation.
        Temperature of the mixing chamber as a function of time during a full cooldown and warm-up cycle.
        Without microwave measurement setup installed, the precooling step (P) brings the system from room temperature to \qty{4}{\kelvin} in approximately \qty{15}{\minute}.
        The dilution step reaches a base temperature of \qty{70}{\milli\kelvin} within \qtyrange{25}{40}{\minute} depending on the thermal load.
        The total turnaround time is about \qty{1.2}{\hour} and up to \qty{2.1}{\hour} with a mounted sample and RF wiring.
    }
    \label{fig:2_cooldown}
\end{figure}

Installing a sample together with its RF environment introduces additional thermal loads that modify these dynamics. The increased heat capacity extends the precooling step to roughly \qty{70}{\minute} and the condensation step to about \qty{40}{\minute}, while the base temperature still stabilizes just below \qty{100}{\milli\kelvin}. In this fully wired configuration, the complete cycle---including measurement and warm-up back to room temperature---takes approximately \qty{2.1}{\hour} (\figref{fig:2_cooldown}).

\subsection{RF measurement setup}

The cryostat is designed to support standard dispersive readout and control of superconducting qubits within its compact geometry. Microwave signals are delivered to the sample through coaxial lines anchored at successive temperature stages, with attenuators distributed along the input line to progressively thermalize the incoming radiation and reduce the effective noise temperature seen by the qubit. Attenuation values are \qty{20}{\deci\bel} at the \qty{4}{\kelvin} stage and \qty{30}{\deci\bel} at the mixing-chamber stage. The sample is mounted and thermally anchored at the mixing chamber stage. Details of the wiring are provided in Appendix \ref{appendix:wiring}.

On the output side, a cryogenic isolator provides at least \qty{20}{\deci\bel} of isolation from amplifier back-action over the \qtyrange{4}{8}{\giga\hertz} band, and a high-electron-mobility-transistor (HEMT) amplifier at the \qty{4}{\kelvin} stage provides the first amplification stage, with a gain of approximately \qty{20}{\deci\bel} and a noise temperature of about \qty{3}{\kelvin} when biased at lower-than-nominal power ; further amplification is performed at room temperature. Calibrating the total gain against known input signals, we estimate a system noise temperature of approximately \qty{10}{\kelvin}, limited by the \qty{20}{\decibel} gain of the HEMT anchored at \qty{4}{\kelvin} and losses between the output of the circulator and the input of the HEMT. Line attenuation is verified independently through resonator spectroscopy and power-dependent measurements. Compared with conventional systems, the compact geometry limits the number of filtering stages that can be accommodated; as discussed below, this constrains the achievable base temperature and coherence but leaves the platform fully capable of qubit characterization.

\section{Fluxonium Qubits Experiment}

\subsection{Two-qubit device}

We demonstrate the capabilities of the cryostat using a chip that hosts two fluxonium qubits, labeled Q1 and Q2, coupled through a microwave resonator. Additionally, each qubit is connected to a dedicated $\lambda/2$ resonator, enabling multiplexed readout through a common feedline. Fluxonium qubits are an especially demanding probe of cryogenic and microwave conditions: their low transition frequency makes them particularly sensitive to thermal effects, their broadband spectrum makes them sensitive to imperfect filtering, and their strong flux dispersion makes them sensitive to flux noise. Successfully operating fluxonium qubits therefore constitutes a stringent test of our cryogenic platform. The qubits are addressable through one charge line for Q1 and one on-chip flux line for Q2, as shown in \figref{fig:3_fluxonium}(a) and (b). An external coil is used to apply a global magnetic field on the device. The chip is mounted in a home-made four-port sample holder at the mixing-chamber stage and connected to the RF measurement chain. Thanks to the rapid cooldown, measurements begin approximately \qty{1.5}{\hour} after sample installation.

\begin{table*}[ht]
\centering
\caption{Summary of device parameters and Hamiltonian fit results. The fluxonium energy scales $E_J$, $E_C$, $E_L$, the external flux bias $\phi_\mathrm{ext}/\phi_0$, the qubit frequencies $\omega_{01}/2\pi$, the readout frequency $\omega_\mathrm{RO}/2\pi$, the qubit--resonator coupling $g$, the resonator linewidth $\kappa$, the qubit lifetime $T_1$, the echo coherence time $T_2^E$, the single-qubit gate duration $t_\mathrm{gate}$, the average Clifford and physical single-qubit gate fidelities $F_\mathrm{Clifford}$ and $F_\mathrm{1QB}$, and the corresponding coherence-limited bound $F_\mathrm{1QB}^\mathrm{max,coh}$ are reported.}
\label{tab:qubit_params}
\resizebox{\textwidth}{!}{%
\begin{tabular}{lcccccccccccccc}
\hline\hline
& $E_J$ (\si{\giga\hertz}) & $E_C$ (\si{\giga\hertz}) & $E_L$ (\si{\giga\hertz}) & $\phi_\mathrm{ext}/\phi_0$ & $\omega_{01}/2\pi$ (\si{\giga\hertz})& $\omega_\mathrm{RO}/2\pi$ (\si{\giga\hertz})& $g$ (\si{\mega\hertz}) & $\kappa/2\pi$ (\si{\mega\hertz}) & $T_1$ (\si{\micro\second}) & $T_2^E$ (\si{\micro\second}) & $t_\mathrm{gate}$ (\si{\nano\second})& $F_\mathrm{Clifford}$ (\si{\percent})& $F_\mathrm{1QB}$ (\si{\percent}) & $F_\mathrm{1QB}^\mathrm{max, coh}$ (\si{\percent}) \\
\hline
Q1 & 1.621 & 1.903 & 0.733 & 0.5 & 2.222 & 6.032 & 195 & 12 & 1.81 & 0.631 & 16 & 98.18 & 99.03 & 99.04 \\
Q2 & 3.317 & 1.763 & 0.525 & 0.5 & 0.962 & 6.858 & 336 & 11 & 2.35 & 0.718 & 16 & 96.39 & 98.07 & 99.13 \\
\hline\hline
\end{tabular}%
}
\end{table*}

\subsection{Spectroscopy and Hamiltonian Extraction}

\begin{figure*}[t]
    \centering
    \includegraphics[width=\linewidth]{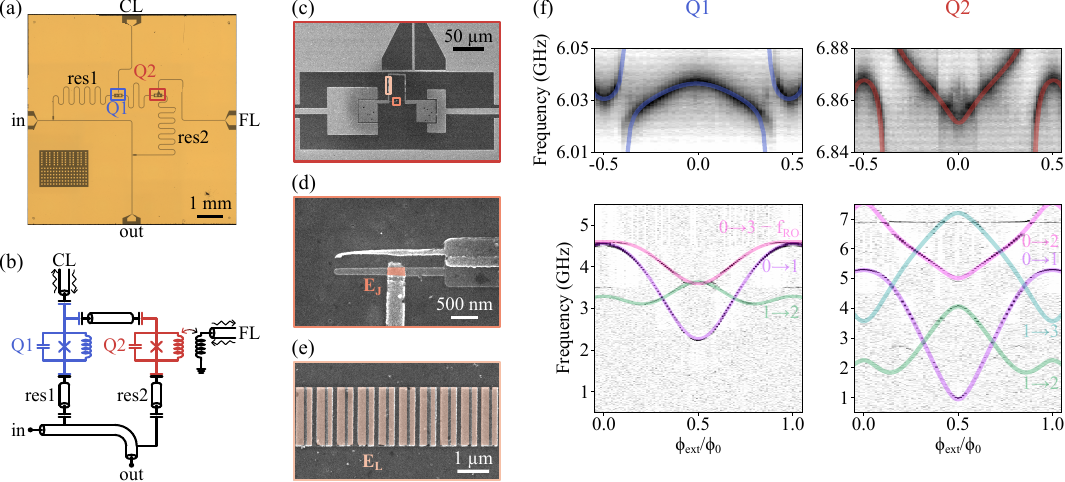}
    \caption{
        Two-fluxonium device characterization.
        (a) Micrograph and (b) schematic of the characterized device.
        (c) SEM image of Q2, with zoom-ins on the Josephson junction (d) and the superinductor (e).
        (f) One-tone spectroscopy of the readout resonator as a function of applied magnetic flux, together with two-tone spectroscopy revealing the fluxonium transition frequencies. Fitting both the resonator and qubit dispersions yields the energy parameters reported in Table~\ref{tab:qubit_params}.
    }
    \label{fig:3_fluxonium}
\end{figure*}

Each fluxonium qubit is described by the Hamiltonian~\cite{doi:10.1126/science.1175552}
\begin{equation}
H = 4 \, E_C \, \hat{n}^2 + \frac{1}{2} \, E_L \left( \hat{\phi} + \phi_\mathrm{ext} \right)^2 - E_J \, \cos{\!\left( \hat{\phi} \right)},
\end{equation}
where $\hat{n}$ and $\hat{\phi}$ are conjugate charge and phase operators. We first use one-tone spectroscopy to identify the bare resonator frequency and its flux dispersion, sweeping the applied magnetic flux over more than one flux quantum to map the periodic modulation of the spectrum around the resonator frequency. We then perform two-tone spectroscopy by applying a weak probe tone at the resonator frequency while sweeping a second drive tone; the resulting transmission signal reveals the qubit transition frequencies, as shown in \figref{fig:3_fluxonium}(f).

The flux dispersions of both the qubit and resonator lines are fitted by numerical diagonalization~\cite{Chitta_2022, Groszkowski2021scqubitspython} of the Hamiltonian, minimizing the deviation between the measured transition frequencies and the theoretical predictions. This procedure yields the energy scales $E_C$, $E_L$, and $E_J$ of each qubit, together with the readout frequency $\omega_\mathrm{RO}/2\pi$ and the qubit--resonator coupling rate $g$. The extracted parameters are reported in Table~\ref{tab:qubit_params}, demonstrating that our ultrafast cryogenic system enables the full parameter extraction of a multi-qubit chip.

\subsection{Time-Domain Characterization and Qubit Performance}

\begin{figure*}[t]
    \centering
    \includegraphics[width=\linewidth]{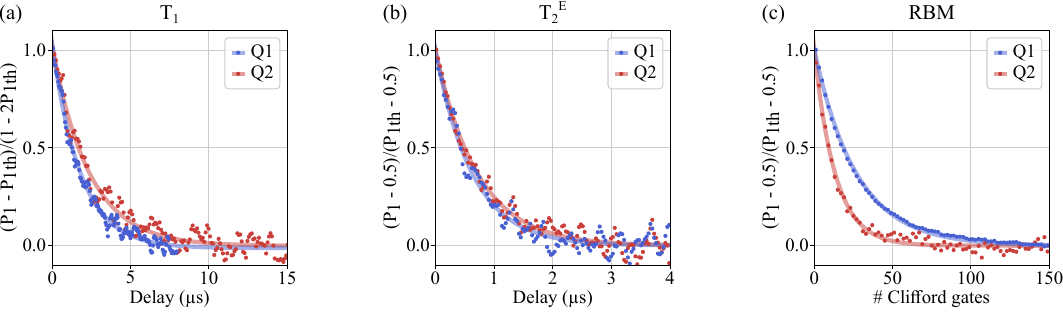}
    \caption{
        Qubit lifetime and randomized benchmarking of Q1 (blue) and Q2 (red).
        (a) Time-domain measurement of energy relaxation showing $T_1 = \qty{1.81}{\micro\second}$ and \qty{2.35}{\micro\second}, respectively.
        (b) Hahn-echo coherence time measurement showing $T_2^E = \qty{0.631}{\micro\second}$ and \qty{0.718}{\micro\second}, respectively.
        (c) Single-qubit randomized benchmarking showing average physical gate fidelity of \qty{99.03}{\percent} and \qty{98.07}{\percent}, respectively.
    }
    \label{fig:4_charac}
\end{figure*}

Time-domain measurements are performed using standard pulsed-microwave techniques on the $0 \!\rightarrow\! 1$ qubit transition of \figref{fig:3_fluxonium}. Each qubit is parked at half flux, where it is first-order insensitive to flux noise; there the transition frequencies are $\omega_{01}/2\pi = \qty{2.222}{\giga\hertz}$ for Q1 and \qty{0.962}{\giga\hertz} for Q2. Qubit excitation uses \qty{16}{\nano\second} Gaussian pulses resonant with this transition; the large anharmonicity of fluxonium makes any additional pulse shaping unnecessary. After calibrating the Rabi frequency, we measure energy relaxation by applying a $\pi$ pulse and recording the decay of the excited-state population as a function of delay time. Fitting the decay to a single exponential (\figref{fig:4_charac}) yields $T_1 = \qty{1.81}{\micro\second}$ for Q1 and \qty{2.35}{\micro\second} for Q2, limited by the base temperature of the mixing chamber. Dephasing is characterized with a Hahn-echo sequence, whose exponential envelope gives $T_2^E = \qty{0.631}{\micro\second}$ for Q1 and \qty{0.718}{\micro\second} for Q2, most likely limited by imperfect filtering of the RF lines.

We next benchmark single-qubit control. We perform single-qubit Clifford randomized benchmarking~\cite{PhysRevA.85.042311}, compiling each of the \num{24} Clifford operations with an XY decomposition that requires on average \num{1.875} physical gates per Clifford~\cite{barends2014logic}. For each sequence depth up to \num{150} gates, we apply \num{200} random Clifford sequences. The average gate fidelity is extracted from a power-law fit $A + B p^m$ of the survival probability, where $p$ is the depolarization parameter, $m$ the number of applied Clifford gates, and $A$ and $B$ absorb state-preparation-and-measurement (SPAM) errors. We obtain average Clifford fidelities of $F_\mathrm{Clifford} = \qty{98.18}{\percent}$ for Q1 and \qty{96.39}{\percent} for Q2, corresponding to average physical single-qubit gate fidelities of $F_\mathrm{1QB} = \qty{99.03}{\percent}$ (Q1) and \qty{98.07}{\percent} (Q2). For reference, Table~\ref{tab:qubit_params} lists the coherence-limited bound $F_\mathrm{1QB}^\mathrm{max,coh}$ computed from the gate duration and the measured coherence times~\cite{Abad2025impactofdecoherence}. Notably, the Q1 gate fidelity (\qty{99.03}{\percent}) coincides with its coherence-limited bound (\qty{99.04}{\percent}), indicating that our single-qubit control is limited by the qubit coherence, itself set by the elevated operating temperature of $\qty{70}{\milli\kelvin}$ and the imperfect filtering of the lines.

The compact geometry constrains filtering and thermalization, which directly shape the noise environment and, in turn, the relaxation times. At a base temperature of \qty{70}{\milli\kelvin}, the thermal occupation is non-negligible for the low-frequency fluxonium transitions and sets an upper bound on the achievable $T_1$. A standard dielectric-loss analysis (Appendix~\ref{appendix:dielectric_loss}) shows that the measured $T_1$ values are consistent with an effective bath temperature $T_\mathrm{eff}^\mathrm{bath} = \qty{177}{\milli\kelvin}$---above the mixing-chamber temperature, as routinely measured for superconducting qubits---and a dielectric quality factor of order \num{8e4}, a standard value for fluxonium qubits. Importantly, our analysis does not show that dielectric loss is the only dissipation mechanism. Other mechanisms—--such as energy relaxation induced by quasiparticle tunneling or thermally-enhanced Purcell effect-—-can contribute significantly to the total observed dissipation. Reassuringly, comparable $T_1$ values have been reported for fluxonium qubits at similar mixing-chamber temperatures and frequencies~\cite{ateshian2025temperaturemagneticfielddependenceenergy}, owing to the rapid increase of dielectric loss with temperature and the multilevel nature of relaxation in fluxonium~\cite{azar2026characterization}.

The measured coherence can be improved substantially by addressing each of these constraints. Lowering the mixing-chamber temperature would reduce thermal occupation; increasing the total input-line attenuation would lower the photon temperature; and adding infrared filters~\cite{serniak2018hot} and a light-tight, radiation-shielded sample holder would suppress quasiparticle generation~\cite{corcoles2011protecting, barends2011minimizing}. Reducing flux noise using a magnetic shield (we presently use a single magnetic shield placed outside the cryostat) and adding cryogenic circulators (only a single isolator was used here, see Appendix \ref{appendix:wiring}) to suppress amplifier back-action would further improve performance. None of these mitigations is fundamentally precluded by the compact architecture; they define a clear path toward higher coherence in future iterations.

\section{Discussion}

The value of the platform is best appreciated by comparing it quantitatively with a conventional commercial dilution refrigerator, as summarized in Table~\ref{tab:comparison}. The most striking difference is turnaround time: a full wired cycle takes \qty{2.1}{\hour}, compared with roughly \qty{48}{\hour} for a conventional wired system---a nearly twentyfold reduction. The footprint is reduced comparably, from a \qty{400}{\milli\meter}-diameter envelope to \qty{100}{\milli\meter}, corresponding to more than an order-of-magnitude smaller cross-sectional area, at a total mass of only \qty{3}{\kilo\gram}.

These gains come with a well-defined trade-off in cryogenic performance. The cooling power at the mixing chamber, \qty{20}{\micro\watt} at \qty{100}{\milli\kelvin}, is about ten times smaller than that of a standard system ($\sim\qty{200}{\micro\watt}$ at \qty{100}{\milli\kelvin}), and the base temperature of our first prototype is higher (\qty{70}{\milli\kelvin} versus \qty{20}{\milli\kelvin}). This reduced cooling budget, together with the limited number of filtering stages that the compact geometry can accommodate, is the primary origin of the elevated effective bath temperature limiting our qubit performances. On the readout side, the same size constraints lead to a system noise temperature of approximately \qty{10}{\kelvin}, set mainly by the \qty{20}{\deci\bel} gain of the \qty{4}{\kelvin} HEMT and microwave losses between the device and this cryoamplifier; this is higher than in setups that can accommodate a quantum-limited parametric amplifier, but remains fully sufficient for the dispersive readout demonstrated here.

Crucially, none of these trade-offs prevented a complete device characterization. We extracted the full circuit Hamiltonian of a two-fluxonium chip, measured its coherence, and reached high single-qubit gate fidelities. The platform is therefore not intended to replace high-performance cryostats for long-coherence experiments, but to complement them by enabling high-throughput screening and parameter extraction with a turnaround time that is otherwise inaccessible.

\begin{table}[t]
\centering
\caption{Comparison of the ultracompact cryostat with a representative conventional large-scale dilution refrigerator. Values for the conventional system are typical figures for a wired qubit-measurement setup.
}
\label{tab:comparison}
\begin{tabular}{lccc}
\hline\hline
 & This work & Conventional & Bottom loader\\
\hline
Base temperature & \qty{70}{\milli\kelvin} & \qty{10}{\milli\kelvin} & \qty{10}{\milli\kelvin} \\
Cooling power at \qty{100}{\milli\kelvin} & \qty{20}{\micro\watt} & \qty{200}{\micro\watt} & \qty{45}{\micro\watt} \\
$^3$He quantity & \qty{1}{\liter} & \qty{10}{\liter} & \qty{10}{\liter} \\
Cooldown time (wired) & \qty{2.1}{\hour} & $\sim\qty{48}{\hour}$ & $\sim\qty{7}{\hour}$ \\
Useful diameter & \qty{52}{\milli\meter} & \qty{300}{\milli\meter} & \qty{60}{\milli\meter} \\
Outer diameter & \qty{100}{\milli\meter} & \qty{400}{\milli\meter} & \qty{400}{\milli\meter} \\
\hline\hline
\end{tabular}
\end{table}

Scaling the platform toward larger multi-qubit experiments is constrained primarily by the available cooling power and the wiring density that the compact geometry can support, which together limit the number of simultaneously addressable qubits and the tolerable thermal load. Several strategies can relax these constraints: frequency-multiplexed readout reduces the number of output lines required, on-chip filtering and impedance engineering reduce the need for bulky cryogenic components, and superconducting parametric amplifiers can improve the readout signal-to-noise ratio. Because each unit is small and inexpensive to operate, the architecture is also naturally compatible with modular deployment of multiple cryostats in parallel, enabling high-throughput screening of quantum devices and accelerating fabrication-feedback cycles.

\section{Conclusion}

We have demonstrated an ultracompact, fast-cycling dilution refrigerator and validated it through the complete characterization of a two-fluxonium device---from full Hamiltonian extraction to single-qubit randomized benchmarking---with a wired turnaround time of \qty{2.1}{\hour}, nearly twenty times faster than a conventional system. Despite a base temperature of \qty{70}{\milli\kelvin} and a modest cooling power of \qty{20}{\micro\watt} at \qty{100}{\milli\kelvin}, the platform supports coherent control at the coherence limit, establishing compact dilution refrigeration as a practical route to high-throughput quantum device characterization.

Future work will focus on lowering the base temperature and improving the electromagnetic shielding and filtering of the RF lines, which together should raise the achievable coherence toward the levels obtained in large-scale systems. We also note that only a small fraction of the \qty{5}{\liter} mixture is actually liquefied, the majority remaining in the gaseous phase in the external gas-handling system; optimizing this balance, together with further miniaturization of the cryogenic RF components, offers additional room for improvement. Beyond research use, the compactness, low operating cost, and short cycle time of the platform make it an attractive tool for education, providing a practical entry point for hands-on training in superconducting quantum technologies.

\section*{Acknowledgments}

This work is supported by the Horizon Europe programme \mbox{HORIZON-CL4-2022-QUANTUM-01-SGA} through the project \mbox{101113946 OpenSuperQPlus100}, by the French grants \mbox{ANR-22-PETQ-0001}, \mbox{0002} and \mbox{0003}, and by \mbox{CRYONEXT} programme under the ``France 2030'' plan.

We thank Prof.~Benjamin Huard for challenging us to demonstrate that complete qubit characterization and high-fidelity control were possible in this cryogenic system. We also thank all members of the Quanteca team at the Institut N\'eel for stimulating discussions and continuous support throughout this work.

\appendix

\section{Wiring of the fridge \label{appendix:wiring}}

\begin{figure}[h]
    \centering
    \includegraphics[width=\linewidth]{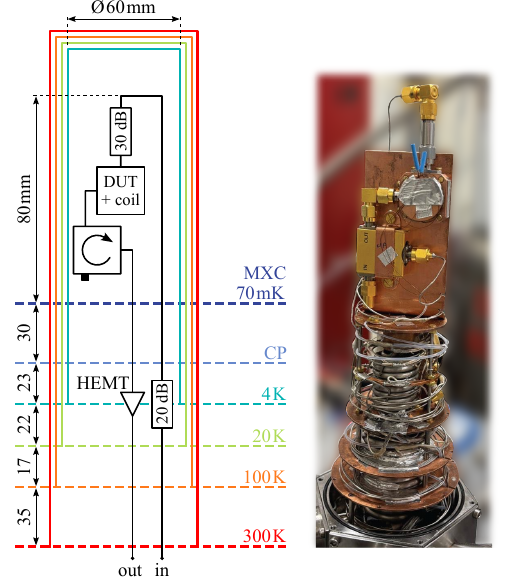}
    \caption{Schematic and photograph of the wiring of the dilution refrigerator.
    }
    \label{fig:5_wiring}
\end{figure}

Figure~\ref{fig:5_wiring} shows the wiring of the dilution refrigerator together with a photograph of the wired system. Because of the space constraints, we use an SMP-to-SMP \qty{20}{\deci\bel} attenuator at the \qty{4}{\kelvin} stage on the input line, a single LNF \qtyrange{4}{8}{\giga\hertz} circulator at the mixing chamber, and a compact cryogenic amplifier (LNF-LNC4\_8SG). Temperature is monitored with home-made, in-house-calibrated ruthenium-oxide sensors, and the coil is driven through NbTi DC cables running from room temperature (not shown in the wiring diagram). Radiation shields are installed on the \qty{100}{\kelvin}, \qty{20}{\kelvin}, and \qty{4}{\kelvin} stages, and an external magnetic shield screens the Earth's magnetic field.

\section{Dielectric Loss \label{appendix:dielectric_loss}}

We use the standard dielectric-loss model for superconducting circuits. The qubit $T_1$ limitation due to capacitive loss can be estimated as~\cite{Nguyen.2019,sun_characterization_2023}
\begin{equation}
\label{eq:capacitiveloss}
\frac{1}{T_\mathrm{1, diel}} = \frac{ \hbar \, \omega_{01}^2 }{ 4 \, E_\mathrm{C} \, Q_\mathrm{cap}(\omega_{01}) } \, \coth{\!\left( \frac{ \hbar \, \omega_{01} }{ 2 \, k_\mathrm{B} \, T_\mathrm{eff}^\mathrm{bath}} \right)} \left| \bra{0}\varphi\ket{1} \right|^2,
\end{equation}
where $\omega_{01}$ is the qubit angular frequency, $Q_\mathrm{cap}(\omega) = Q_\mathrm{cap}(\omega_\mathrm{r})(\omega_\mathrm{r}/\omega)^\epsilon$ is the quality factor associated with capacitive loss, and $T_\mathrm{eff}^\mathrm{bath}$ is the temperature of a bath of weakly coupled two-level systems. We evaluate this expression with $\epsilon = 0.2$ and $\omega_\mathrm{r} = 2 \pi \times \qty{6}{\giga\hertz}$.

Fitting the measured $T_1$ values of Table~\ref{tab:qubit_params} to Eq.~\eqref{eq:capacitiveloss}, we obtain $T_\mathrm{eff}^\mathrm{bath} = \qty{177}{\milli\kelvin}$ and $Q_\mathrm{cap}(\omega_\mathrm{r}) \approx \num{8.2e4}$, both typical for this loss channel given the elevated operating temperature of the cryostat. The model then predicts $T_1 = \qty{1.75}{\micro\second}$ for Q1 and $\qty{2.34}{\micro\second}$ for Q2, in good agreement with the measured values. We further note that a very similar $Q_\mathrm{cap}(\omega_\mathrm{r})$ was extracted in Ref.~\cite{PhysRevX.14.041014} using the same fabrication recipe.

\bibliography{biblio}

\end{document}